\documentclass{elsart}
\usepackage{graphicx}
\usepackage{amssymb}

\begin{document}

\begin{frontmatter}

\title{Fluctuation Theorem in Rachet System}

\author{T. Monnai}
\ead{monnai@suou.waseda.jp}

\address{Department of Applied Physics ,Waseda University,3-4-1 Okubo,
Shinjuku-ku,Tokyo 169-8555,Japan }

\begin{abstract}
Fluctuation Theorem(FT) has been studied as far from equilibrium theorem, which relates 
the symmetry of entropy production.
To investigate the application of this theorem, especially to biological physics, we 
 consider the  FT for tilted rachet system.
Under, natural assumption, FT for steady state is derived. 
\end{abstract}

\end{frontmatter}

\section{INTRODUCTION}
Since the discovery by Evans.et.al.\cite{EvansCohen}, Fluctuation Theorem(FT) has been studied in many
 situations, both stochastic\cite{Crooks},\cite{Kurchan},\cite{Spohn1999} and deterministic\cite{EvansCohen}\cite{Jarzynski}, though systems and interpretations differ, 
FT has a universal form :
\begin{equation}
\frac{Prob(\Delta S=A)}{Prob(\Delta S=-A)}\simeq e^{\Delta S}
\end{equation}
here, $\Delta S$ is the entropy generated, and $Prob(\Delta S =A)$ is the probability for $\Delta S =A$.
It is interesting that FT kind relation  universally holds, but its direct experimental application
 is not known.(Though Jarzynski equality\cite{JarzynskiEq} is closely related to FT and has been used in the experiment, this equality is not FT itself, in this sense application of Jarzynsky equality is not direct.) 
So, we would like to derive FT for rachet system, because it relates biological physics, and direct application of FT is expected.
The analysis is based on Langevin treatment.
Kurchan showed that for Langevin system, if the initial state is Gibbsian, FT holds generally\cite{Kurchan}.
Though we consider Langevin treatment, the result does not depend on the initial condition and in this sense our interest is different from that of Kurchan. 

\section{RACHET SYSTEM}

In this paper we consider the rachet system, namely a particle on a periodic potential $V(x+L)=V(x)$ is dragged 
by a constant load force $F$.
Then, the effective potential the particle feels is $V_{eff}\equiv V(x)-x F$.
If, the potential barrier is high enough, the particle can be considerd to be almost at equilibrium and may satisfy the detailed balance relation.
Indeed, the quantitative argument is given in the review by Reimann\cite{ReimannP}.
We summarize the work by Reimann.

We consider a system that a particle on a periodic potential(period $L$)
 is dragged by a constant load force $F$.
The motion of this particle can be described by the Langevin eq. :
\begin{eqnarray}
\eta\frac{d}{dt}x(t)=&&-\frac{d}{dx}V_{eff}(x)+\xi(t) \nonumber \\
V_{eff}(x)\equiv&&V(x)-F x \nonumber \\
\left\langle\xi(t) \xi(s)\right\rangle =&&\sqrt{2 \eta k_B T}\delta(t-s) \nonumber
\end{eqnarray}
The corresponding Fokker-Planck equation becomes :

\begin{equation}
\frac{\partial}{\partial t} P(x,t) =
\frac{\partial}{\partial x} (\frac{dV_{eff}}{d x}\frac{1}{\eta}+\frac{k_B T}{\eta}\frac{\partial}{\partial x} ) P(x,t)= -\frac{\partial }{\partial  x}J(x,t)
\end{equation}   
Here, $J(x,t)$ is the probability current.
Then at a steady state, the reduced probability current :
\begin{equation}
\hat{J}(x,t)\equiv \Sigma_{n=-\infty}^{\infty}J(x+n L,t)  
\end{equation}
becomes :
\begin{eqnarray}
\hat{J}^{st}=&&N (1-e^{-\frac{F L}{k_B T}}) \nonumber \\
N=&&\frac{k_B T}{\eta}(\int_0^L dx \int_x^{x+L}dy e^{\frac{V_{eff}(y)-V_{eff}(x)}{k_B T}})^{-1}
\end{eqnarray}
If in each period $L$ there exists one minimum $x_{min}$ and one maximum $x_{max}$, and the system
is in the weak noise regime 
$k_B T\ll \Delta V_{eff}\equiv V_{eff}(x_{max})-V_{eff}(x_{min})$,
 the saddle point approximation gives the following probability
 current :
\begin{eqnarray}
\hat{J}^{st}=&&k_+-k_- \nonumber \\
k_+=&&\frac{\mid \frac{d^2}{d x^2}V_{eff}(x_{max})\frac{d^2}{d x^2}V_{eff}(x_{min})\mid^{\frac{1}{2}}}
{2 \pi \eta}e^{-\frac{\Delta V_{eff}}{k_B T}} \nonumber \\
k_- =&& k_+ e^{-\frac{F L}{k_B T}}
\end{eqnarray}
Here, $k_+$ and $k_-$ are the Kramers escape rate, the probability per unit time for a particle near a potential minimum escape 
to the right minimum and left minimum respectively.  These $k_+$ and
$k_-$ satisfy the detaled balance relation.
\begin{equation}
\frac{k_+}{k_-}=e^{\frac{F L}{k_B T}}\label{ratio}
\end{equation}
\begin{figure}
\center{
\includegraphics[scale=0.8]{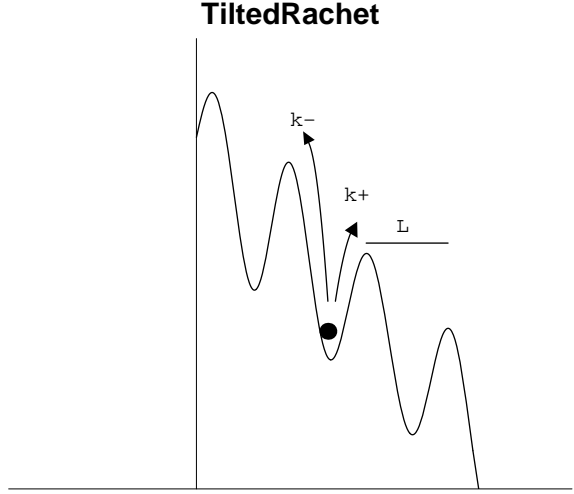}
}
\caption{a particle hopping on tilted potential }
\label{Figg}
\end{figure}
\section{DERIVATION OF STEADY STATE FLUCTUATION THEOREM}
Because of the assumption $k_B T\ll \Delta V_{eff}$, we can describe the 
motion of a particle in terms of biased random walk.
Namely, it is sufficient to observe only the local minmum of effective potential and on what minimum
the particle (approximately) exists at each discrete time $t=0,\tau, 2\tau,...$.
It is natural to assume that a particle hops only to the nearest-neighbor local minima :
to the right minimum with probability $k_+$ 
, to the left minimum with $k_-$, and it does not move with probability $1-(k_++k_-)$.

Then the motion of a particle follows the biased random walk.
A particle hops to right, left, and remains with probabilities $k_+$, $k_-$, and $1-(k_++k_-)$.
\begin{figure}
\center{
\includegraphics[scale=0.8]{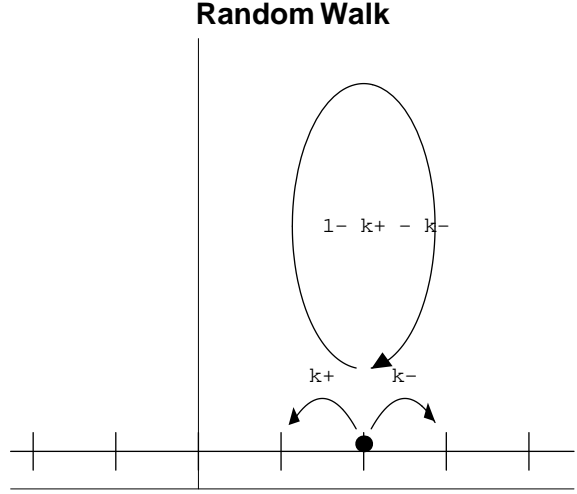}
}
\caption{biased random walk of a particle per unit time}
\label{Figg}
\end{figure}
Next we consider the probability $\pi_n(k)$ which is the probability that after n unit time, 
particle moves k toward right.
Here, we can show the equation:
\begin{equation}
\frac{\pi_n(k)}{\pi_n(-k)}=(\frac{k_+}{k_-})^k
\end{equation}

One can prove this equation via following one to one correspondence.:
Consider any process that causes the movement k after n unit time.
The process is constituted by movements : $k+l$ times toward right, $l$ times toward left, and $n-k-2l$
 times remaining.
This process occurs with probability 
\begin{equation}
\frac{n!}{(k+l)!l!(n-k-2l)!}k_+^{k+l}k_-^l(1-k_+-k_-)^{n-k-2l}
\end{equation}
The corresponding opposite process : $l$ times toward right, $k+l$ times toward left, and $n-k-2l$ times 
remaining.
This process occurs with probability
\begin{equation}
\frac{n!}{(k+l)!l!(n-k-2l)!}k_+^lk_-^{k+l}(1-k_+-k_-)^{n-k-2l}
\end{equation}
So, the ratio of the probability that these two opposite processes occurs is $(\frac{k_+}{k_-})^k$.
This is valid for any $l$, and the equation is derived.

Then we substitute the detaled balance relation for Kramers escape rate $\frac{k_+}{k_-}=e^{\frac{F L}{k_B T}}$
 into the above equation, we obtain following expression:
\begin{eqnarray}
\frac{\pi_n(k)}{\pi_n(-k)}=e^{\frac{F L k}{k_B T}}
\end{eqnarray}
We define the entropy generated $\Delta S\equiv \frac{\Delta Q}{k_B T}$, where $\Delta Q=F L k$ is the 
Joule heat.
$\Delta S$ can be considered as the entropy generated, since the work done on the system
extenally will be entirely dissipated, and there are no contribution to the internal energy. 
Finally, we take the continuous limit, $n\rightarrow t, L k\rightarrow x$, we obtain the steady state
Fluctuation Theorem:
\begin{equation}
\frac{\pi_t(\Delta S)}{\pi_t(-\Delta S)}=e^{\Delta S}
\end{equation}
\section{DISCUSSION}
FT for rachet system is derived.
FT derived here is applicable to any system which obeys a biased random walk and satisfies the detailed balance relation.

Especially, a complete example was recently given by Nishiyama.et.al\cite{Higuchi}.
They performed an experiment where a single kinesin externally forced moves along a microtuble obeying  
a biased random walk with regular $8nm$ steps. And they measured the ratio of the forward to backward movements at each load force
 $1\sim 9pN$, the result well agrees with the detailed balance relation. FT derived in this letter would
 therefore valid for this system. 

Because the derivation of FT consists of two parts, identity (\ref{ratio}) and detailed balance relation,
 there is the possibility that under some condition detailed balance does not hold and FT should be modified.
This modification of FT is an open problem.
\section{Acknowledgement}
The author is grateful to Professor P.Gaspard, and Professor S.Tasaki for fruitful discussions.

\end{document}